\documentclass[preprint,aps,showpacs]{revtex4}

\usepackage{makeidx}
\usepackage{graphicx}

\begin{document}

\title{First-passage and risk evaluation under stochastic volatility} 

\author{Jaume Masoliver}
\email{jaume.masoliver@ub.edu}
\affiliation{Departament de F\'{\i}sica Fonamental, Universitat de Barcelona,\\
Diagonal, 647, E-08028 Barcelona, Spain}
\author{Josep Perell\'o}
\email{josep.perello@ub.edu}
\affiliation{Departament de F\'{\i}sica Fonamental, Universitat de Barcelona,\\
Diagonal, 647, E-08028 Barcelona, Spain}

\date{\today}

\begin{abstract}

We solve the first-passage problem for the Heston random diffusion model. We obtain exact analytical expressions for the survival and hitting probabilities to a given  level of return. We study several asymptotic behaviors and obtain approximate forms of these probabilities  which prove, among other interesting properties, the non-existence of a mean first-passage time. One significant result is the evidence of extreme deviations --which implies a high risk of default-- when certain dimensionless parameter, related to the strength of the volatility fluctuations, increases. We believe that this may provide an effective tool for risk control which can be readily applicable to real markets.

\end{abstract}
\maketitle

\section{Introduction}
\label{sec1}

The study of first-passage and exit problems have a long and standing tradition in physics, engineering and natural sciences \cite{george_book,redner}. Perhaps the most important example of an exit problem in physics is provided by the ``Kramers problem'' where one studies the possible escape, owing to noise, of a system from a stable estate \cite{weiss_86,hanggi}. Classical examples of first-passage problems are the collapse of mechanical structures because of random external vibrations which attain an extreme amplitude beyond the stability threshold; or the ``false alarm problem'' where internal fluctuations induce the current or voltage of an electric circuit to reach a critical value for which an alarm is triggered \cite{roberts,kl}.

In finance, the study of extreme events from the perspective of first-passage or exit problems is still beginning. In few recent papers we have address the escape problem in the context of econophysics using several techniques for different settings \cite{mmp,montero_lillo,mp,mm_2007,met_heston}. In the present paper we want to address for financial series the related issue of the first-passage problem. This is a relevant question because it is intimately connected to extreme values and, therefore, to risk and default. 

Needless to say that the evaluation and subsequent risk control should be central issues in finance. Crisis are, nevertheless, inherent characteristics of financial markets although one expects them to be rare. Unfortunately this trend is changing, at least in the last two decades, for the frequency of appearance of extreme events seems to be increasing and there is a  growing consensus that there should be a change of model, specially in the fields of option pricing and structured products \cite{bouchaud}. 

Traditionally tasks related to risk control have been based on the assumption that prices follow the geometric Brownian motion (GBM). However, it is known for long \cite{mandelbrot,fama} that the Gaussian distributions resulting from the GBM decay for large values of the price (or the return) much more steeply than the empirical distributions. In other words, real financial distributions present fat tails and prices are highly leptokurtic \cite{cont}. 

Therefore, the GBM model clearly underestimates risk. Indeed, risk is associated with large deviations of returns and the appearance of these deviations is determined by the tails of the probability distribution. Since for the GBM tails are much thinner than the empirical ones, we conclude that actual risk is higher than the risk foreseen by the GBM. Practitioners usually remedy this (dangerous \cite{bouchaud}) inadequacy of the GBM model by the ad-hoc procedure of assuming that prices follow a mixture of two Gaussian distributions. The mixture is tailored to reproduce the real probability distribution, that is to say, one Gaussian curve adjusts the center of the real distribution and the second one the tails \cite{mixture}. Whereas this certainly improves risk management, it is an unsatisfactory procedure, for it is little more than curve fitting without any model sustaining it. Similar objections can be raised to other approaches based on adjusting actual prices by truncated Levy distributions \cite{mantegna,koponen,matacz}.

Another facet of the problem is provided by the fact that the GBM assumes a constant volatility. The latter defined  as the standard deviation of returns. However, the recording of the empirical prices of financial options clearly indicates that the volatility implied in these prices is not a constant, not even a function of time, but a random variable \cite{hull87,engle-pat}. This gave rise in the late ninety eighties to the so-called stochastic volatility (SV) models for which return and volatility constitute a two-dimensional diffusion process governed by a pair of Langevin equations \cite{IJTAF,QF,fouque_book}. 

Financial models based on SV have been gaining increasing acceptance, not only for the initial motivation of correcting option prices, but because they provide a natural and reasonable explanation to many empirical observations which are gathered together under the collective name of ``stylized facts'' \cite{cont}. In particular SV models result in fat-tailed distributions of returns which overcomes one of the main objections against the GBM. We should also note that volatility is today a crucial concept in any financial setting and many financial products are based on it. Therefore, the risk associated with volatility is particularly important from theoretical, as well as practical, aspects of the problem.  

The measure of risk is provided by the probability of attaining an extreme value (or a preassigned critical label) and, in the standard approach, this probability is obtained through the distribution of prices. In such a procedure the probability of reaching certain critical mark $L$ is secured by counting all times the process has reached the label $L$. There is an alternative (and finer) approach to the problem which consists in counting only the event of reaching $L$ for {\it the first time}. This is the {\it first-passage problem} that, as we have explained above, is an old and well known acquaintance of physics, engineering and natural sciences \cite{george_book,redner}. Within the context of continuous-time random walks, we have recently shown that risk control methods based on first-passage techniques are more efficient than traditional ones, since the latter clearly underestimate risk \cite{mm_2007}.

Herein we address the first-passage (or hitting) problem for financial time series under the assumption of stochastic volatility. We choose one particular SV model, the Heston model, because it has the advantage of allowing exact analytical results \cite{heston,dragulescu,silva,bollerslev}. In a very recent paper \cite{met_heston} we have solved a closely related issue: the escape problem for the Heston model. In that problem the process can exit, for the first time, a given interval and this leads to a two-barrier problem. However, in the first-passage problem the process only crosses a given critical value which amounts to solving an one-barrier problem. 

To our knowledge there are a few but increasing number of works dealing with first-passage problems \cite{simonsen,fouque1,fouque2} and exit problems  \cite{mmp,montero_lillo,mp,mm_2007,met_heston,bonano1,bonano2,spagnolo2} in finance. Many of them address the exit problem in a restricted way, because they only obtain the mean exit time, i.e., the first moment of the exit distribution. This moment is easier to handle than the entire distribution, although it provides much less information, specially about the time evolution of the exit problem. On the other hand first-passage problems are usually more intricate because in many cases the mean first-passage time does not exist. This is the case, for instance, of the Brownian motion or the standard SV models (see below). In such situations one is either bound to get the entire first-passage distribution, which is usually quite involved, or to introduce ad-hoc forces for securing the existence of the first moment of the hitting  \cite{bonano1,bonano2}.  

In this paper we solve the complete first-passage problem for the Heston SV model. We obtain the exact expression of the survival probability and show that the mean first-passage time does not exist. We also average the volatility in order to work out the problem for the return alone. We obtain approximate expressions of the survival probability in several asymptotic regimes --long times, large volatilities and small volatility fluctuations-- and find that in these cases the (approximate) survival probability is Gaussian and has the same form than that of the Wiener process. We also obtain the asymptotic expression for survival when the fluctuations of the volatility are high. In the later case the Gaussian character is lost and the probability of hitting an extreme value is much higher than in the case of mild volatility fluctuations.

The paper in organized as follows. In Sec. \ref{sec2} we present the Heston model and obtain the exact solution to the first-passage problem. In Sec. \ref{sec3} we obtain the approximate expression for the survival probability in several asymptotic limits. In Sec. \ref{sec4} we treat the problem according to the fluctuations of the volatility. In Sec. \ref{sec5} we average out the volatility assuming it has reached the stationary state. This allows us to get exact as well as approximate expressions for the survival probability of the return alone. A brief summary of the main results along with few conclusions are given in Sec. \ref{sec6}. Some more technical details are in an Appendix.

\section{The Heston model and the first-passage time distribution}
\label{sec2}

Let $X(t)$ be the zero-mean return $X(t)$ defined through the stochastic differential (in the It\^o sense):
\begin{equation}
dX(t)=\frac{dP(t)}{P(t)}-\left\langle\frac{dP(t)}{P(t)}\right\rangle,
\label{zero-mean}
\end{equation}
where $P(t)$ is a speculative price or the value of an index and $\langle\cdot\rangle$ denotes the average. In terms of $X(t)$ the Heston model \cite{heston} is a two-dimensional diffusion process $(X(t),Y(t))$ described by the following pair of stochastic differential equations (again, in the It\^o sense)
\begin{equation}
dX(t)=\sqrt{Y(t)}dW_1(t),
\label{dx}
\end{equation}
\begin{equation}
dY(t)=-\alpha\left[Y(t)-m^2\right]dt+k\sqrt{Y(t)}dW_2(t),
\label{dy}
\end{equation}
where $W_i(t)$ are Wiener processes, i.e. $dW_i(t)=\xi_i(t)dt$ $(i=1,2)$, where $\xi_i(t)$ are zero-mean Gaussian white noises with $\langle\xi_i(t)\xi_i(t')\rangle=\delta_{ij}\delta(t-t')$ \cite{footnote1}. 

Equation (\ref{dx}) shows that $Y(t)$ is the variance of return and the volatility $\sigma(t)$ is given by
\begin{equation}
\sigma(t)=\sqrt{Y(t)}.
\label{volatility}
\end{equation}
However, as long as no confusion arises, we will use the term ``volatility variable'' or just ``volatility'' for the random process $Y(t)$. Moreover, as proved many years ago \cite{feller}, the volatility process $Y(t)$ defined by Eq. (\ref{dy}) is positive and the volatility is well defined.  

In Eq. (\ref{dy}) the parameter $m$ is the so-called ``normal level of volatility'',  $\alpha>0$ is related to the ``reverting force'' toward the normal level $m$ (which cannot be zero, see discussion below) and $k$, sometimes referred to as the ``vol of vol'', measures the fluctuations of the volatility. It is useful to keep in mind that the quantities $Y(t)$, $\alpha$, $m^2$, and $k$ have them all units of 1/times; while the zero-mean return and, consequently $L$ are dimensionless. 

One can easily see that $Y(t)$ is a homogeneous and stationary random process whose probability density function as $t\rightarrow\infty$ (or, equivalently as $t_0\rightarrow-\infty$, where $t_0$ is the initial time) is given by the Gamma distribution:
\begin{equation}
p_{st}(y)=\frac{(2\alpha/k^2)^\nu}{\Gamma(\nu)}y^{\nu-1}e^{-(2\alpha/k^2) y},
\label{Gamma}
\end{equation}
where $\nu=(2\alpha/k^2)m^2$.

From Eq. (\ref{Gamma}) we readily see that the stationary mean value and the stationary variance of the volatility variable are  
\begin{equation}
\langle Y(t)\rangle_{\rm st}=m^2 \qquad {\rm Var}\left\{Y(t)\right\}_{\rm st}=(k^2/2\alpha)m^2.
\label{variance}
\end{equation}
As long as $Y(t)$ is a (positive) random variable neither its variance nor its mean can be zero, hence $m\neq 0$. 

After this brief summary on the Heston model we turn our attention to first-passage problems. Let $S(x,y,t)$ be the probability that the zero mean return, $X(t)$, initially at $X(0)=x$ \cite{x=0} with volatility $Y(0)=y$, has never crossed the critical level $L$ before time $t$. 

When $L<x$ (i.e., $L$ is a level of losses) $S(x,y,t)$ coincides with the survival probability (SP) for the joint process $(X(t),Y(t))$ to be, at time $t$, still inside the semi-infinite strip
$$
L<X(t)<\infty, \qquad 0<Y(t)<\infty.
$$
In such a case the (first) hitting probability, that is, the probability of having a loss labeled by $L$ (if $L\rightarrow-\infty$, this is the ``default probability'') is given by
\begin{equation}
W(x,y,t)=1-S(x,y,t).
\label{default}
\end{equation}

When $L>x$, $S(x,y,t)$ is the SP of the semi-infinite strip 
$$
-\infty<X(t)<L, \qquad 0<Y(t)<\infty,
$$
and the probability of hitting a profit $L$ is also given by Eq. (\ref{default}). Since in our Heston model we do not consider any bias both situations --loss and profit-- are symmetrical.

The survival probability $S(x,y,t)$ is the solution to the following initial and boundary value problem \cite{gardiner}
\begin{equation}
\frac{\partial S}{\partial t}=-\alpha(y-m^2)\frac{\partial S}{\partial y}+
\frac{1}{2}k^2y\frac{\partial^2 S}{\partial y^2}+\frac{1}{2}y\frac{\partial^2 S}{\partial x^2},
\label{fpe}
\end{equation}
with initial and boundary conditions respectively given by
\begin{equation}
S(x,y,0)=1,\qquad S(L,y,t)=0.
\label{ibc}
\end{equation}
Note that the symmetry between losses and profits just mentioned can be readily seen by the fact that under the change of variable
\begin{equation}
z=|L-x|
\label{z}
\end{equation}
problem (\ref{fpe})-(\ref{ibc}) remains unchanged:
\begin{equation}
\frac{\partial S}{\partial t}=-\alpha(y-m^2)\frac{\partial S}{\partial y}+
\frac{1}{2}k^2y\frac{\partial^2 S}{\partial y^2}+\frac{1}{2}y\frac{\partial^2 S}{\partial z^2},
\label{fpe_z}
\end{equation}
with
\begin{equation}
S(z,y,0)=1,\qquad {\rm and}\qquad S(0,y,t)=0.
\label{ibc_z}
\end{equation}

Since $z\geq 0$ and $S(0,y,t)=0$, we can define the Fourier transform
\begin{equation}
\tilde{S}(\omega,y,t)=\int_0^\infty S(z,y,t)\sin\omega z dz,
\label{ft}
\end{equation}
which turns Eqs. (\ref{fpe_z})-(\ref{ibc_z}) into the following initial value problem
\begin{equation}
\frac{\partial \tilde{S}}{\partial t}=-\alpha(y-m^2)\frac{\partial \tilde{S}}{\partial y}+
\frac{1}{2}k^2y\frac{\partial^2 \tilde{S}}{\partial y^2}-
\frac{1}{2}\omega^2 y \tilde{S},
\label{fpe1}
\end{equation}
with initial condition
\begin{equation}
\tilde{S}(\omega,y,0)=\mathcal{P}\left(1/\omega\right),
\label{ic}
\end{equation}
where
\begin{equation}
\mathcal{P}(1/\omega)=\int_0^\infty \sin\omega z dz=1/\omega \qquad(\omega\neq 0),
\label{PV}
\end{equation}
is the Cauchy principal value \cite{vladimirov}. 

After the definition of the dimensionless variables
\begin{equation}
\tau=\alpha t, \qquad v=y/\alpha, 
\label{tau-v}
\end{equation}
the problem above reads
\begin{equation}
\frac{\partial \tilde{S}}{\partial\tau}=-(v-\theta)\frac{\partial \tilde{S}}{\partial v}+
(\beta^2 v/2)\frac{\partial^2 \tilde{S}}{\partial v^2}-(\omega^2/2)v\tilde{S},
\label{fpe-new}
\end{equation}
and
\begin{equation}
\tilde{S}(\omega,v,0)=\mathcal{P}(1/\omega),
\label{ic-new}
\end{equation}
where 
\begin{equation}
\theta=m^2/\alpha,
\label{theta}
\end{equation}
is the (dimensionless) normal level of volatility and 
\begin{equation}
\beta\equiv k/\alpha,
\label{beta}
\end{equation}
is a dimensionless parameter which gauges the volatility fluctuations with respect to the deterministic strength,  measured by $\alpha$, toward the normal level. 
 
As can be easily seen by direct substitution the solution to the problem posed by Eqs. (\ref{fpe-new})-(\ref{ic-new}) is furnished by
\begin{equation}
\tilde{S}(\omega,v,\tau)=\mathcal{P}\left(1/\omega\right)\exp\left\{-A(\omega,\tau)-2B(\omega,\tau)v/\beta^2\right\},
\label{fourier-solution_0}
\end{equation}
where
\begin{equation}
A(\omega,\tau)=(2\theta/\beta^2)\int_0^\tau B(\omega,s)ds,
\label{A0}
\end{equation}
and $B(\omega,\tau)$ obeys the Riccati equation
\begin{equation}
\dot{B}=-B-B^2+(\beta\omega/2)^2,
\label{riccati}
\end{equation}
with initial condition $B(\omega,0)=0$. 

The explicit expression for the function $A(\omega,\tau)$ and $B(\omega,\tau)$ are obtained in Appendix \ref{appB}. Thus
\begin{equation}
A(\omega,\tau)=(2\theta/\beta^2)\left\{\mu_-(\omega)\tau+\ln\left[\frac{\mu_+(\omega)+\mu_-(\omega)e^{-\Delta(\omega)\tau}}{\Delta(\omega)}\right]\right\},
\label{A}
\end{equation}
and 
\begin{equation}
B(\omega,\tau)=\mu_-(\omega)
\frac{1-e^{-\Delta(\omega)\tau}}{1+[\mu_-(\omega)/\mu_+(\omega)]e^{-\Delta(\omega)\tau}},
\label{B}
\end{equation}
where
\begin{equation}
\Delta(\omega)=\sqrt{1+(\beta\omega)^2},\qquad\qquad \mu_{\pm}(\omega)=[\Delta(\omega)\pm 1]/2.
\label{delta}
\end{equation}
Hence (cf Eqs. (\ref{fourier-solution_0}), (\ref{A}) and (\ref{B}))
\begin{equation}
\tilde{S}(\omega,v,\tau)=\mathcal{P}(1/\omega)\left[\frac{\Delta(\omega)e^{-\mu_-(\omega)\tau}}
{\mu_+(\omega)+\mu_-(\omega)e^{-\Delta(\omega)\tau}}\right]^{2\theta/\beta^2} e^{-(2/\beta^2)B(\omega,\tau)v}.
\label{fourier-solution}
\end{equation}

The solution to the hitting problem for the two-dimensional Heston SV model is therefore given by the Fourier inversion
$$
S(z,v,\tau)=\frac{2}{\pi}\int_0^\infty\tilde{S}(\omega,v,\tau)\sin\omega z d\omega.
$$
Plugging into this equation the expression for $\tilde{S}(\omega,v,\tau)$ given in Eq. (\ref{fourier-solution}) and taking into account Eq. (\ref{PV}), we get
\begin{equation}
S(z,v,\tau)=\frac{2}{\pi}\int_0^\infty \frac{\sin\omega z}{\omega}
\left[\frac{\Delta(\omega)e^{-\mu_-(\omega)\tau}}{\mu_+(\omega)+\mu_-(\omega)e^{-\Delta(\omega)\tau}}\right]^{2\theta/\beta^2}
e^{-(2/\beta^2)B(\omega,\tau)v}d\omega.
\label{SP}
\end{equation}

\begin{figure}[htb]
\begin{center}
\includegraphics{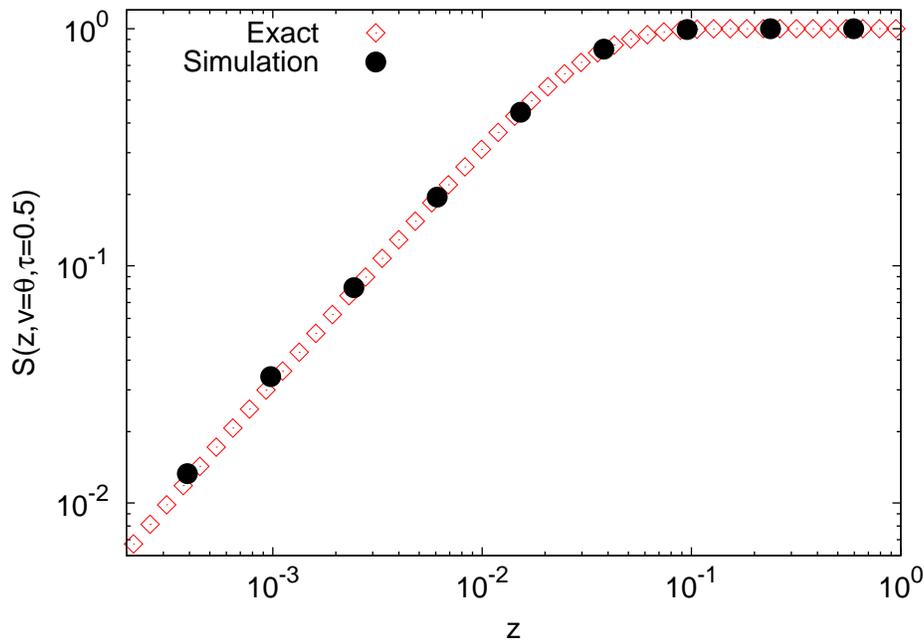}
\caption{(Color online) Log-log representation of the survival probability $S(z,v,\tau)$ as a function of $z$ for $\tau=0.5$ (around 11 trading days), $v=\theta$ and $\beta=0.1$. Empty diamonds represent the numerical evaluation of the exact expression gien in Eq.~(\ref{exact_SP}). Solid circles represent simulation points of the Heston model. We use the realistic parameters values presented in Ref. \cite{dragulescu} which are similar to those encountered in actual markets: $\alpha=0.045$ 1/day, $m^2=8.62\times 10^{-5}$ 1/day and hence $\theta=1.92\times 10^{-3}$.}
\label{fig1}
\end{center}
\end{figure}

In Fig. \ref{fig1} we represent the numerical evaluation of this Fourier integral. We also show there the SP obtained through numerical simulations of the Heston model \cite{moro}. The match between both results is an indication of the correctness of our calculations.

\section{asymptotic approximations}
\label{sec3}

Equation (\ref{SP}) provides the complete solution to the first-passage problem for the Heston model which can be used to get numerical values for the SP and hitting probabilities as is shown in Fig. \ref{fig1} as well as in the rest of figures of the paper. However, it is not the most convenient form of $S(z,v,\tau)$ for bringing into light general properties of the first-passage problem. We will, therefore, try to find approximate expressions that may uncover as many features of the SP as possible. We first obtain asymptotic expressions of the SP for long times which, in turn, will prove the non existence of an average hitting (i.e., first-passage) time. Subsequently we will obtain approximate expressions of the SP valid for large to moderate volatilities. 

We should also note that both approximations --i.e., long times and large volatilities-- coincide. As we will see next this coincidence is due to the particular form of the functions $\mu_-(\omega)$ and $B(\omega,\tau)$ defined above.

\subsection{Long-time asymptotic expressions and mean-first passage time}
\label{subsec31}

We now focus on the behavior of the SP in the asymptotic limit when $t\rightarrow\infty$, to this end rewrite Eq. (\ref{SP}) in the following form:
\begin{equation}
S(z,v,\tau)=\frac{2}{\pi}\int_0^\infty \frac{\sin\omega z}{\omega}
\left[\frac{\Delta(\omega)e^{-B(\omega,\tau)v/\theta}}{\mu_+(\omega)+\mu_-(\omega)e^{-\Delta(\omega)\tau}}\right]^{2\theta/\beta^2} e^{-(2\theta\tau/\beta^2)\mu_-(\omega)}d\omega.
\label{exact_SP}
\end{equation}
For times such that 
$$
(\theta/\beta^2)\tau\gg 1,
$$
we are entitled to use the saddle-point method to secure an approximation of the integral in Eq. (\ref{exact_SP}). One can easily show that $\omega=0$ is a minimum of $\mu_-(\omega)$ (cf. Eq. (\ref{delta})) and expanding around this minimum yields
\begin{equation}
\mu_-(\omega)=(\beta\omega/2)^2+O(\omega^4).
\label{minimum_mu}
\end{equation}
Then
\begin{equation}
S(z,v,\tau)\simeq\frac{2}{\pi}\int_0^\infty \frac{\sin\omega z}{\omega}
\left[\frac{\Delta(\omega)e^{-B(\omega,\tau)v/\theta}}{\mu_+(\omega)+\mu_-(\omega)e^{-\Delta(\omega)\tau}}\right]^{2\theta/\beta^2} e^{-(\theta\tau/\beta^2)\omega^2}d\omega.
\label{S_expan}
\end{equation}

Following the saddle-point method \cite{erdelyi} we expand to low orders in $\omega$ the argument of the exponential term (see Eq. (\ref{B}))
\begin{equation}
B(\omega,\tau)=(\beta\omega/2)^2(1-e^{-\tau})+O(\omega^4).
\label{b_expan}
\end{equation}
Then taking into account Eq. (\ref{minimum_mu}) and (cf. Eq. (\ref{delta}))
\begin{equation}
\Delta(\omega)=1+(\beta\omega)^2/2+O(\omega^4), \qquad \mu_+(\omega)=1+(\beta\omega/2)^2+O(\omega^4),   
\label{delta_expan}
\end{equation}
we have
\begin{equation}
\left[\frac{\Delta(\omega)}
{\mu_+(\omega)+\mu_-(\omega)e^{-\Delta(\omega)\tau}}\right]^{2\theta/\beta^2} e^{-(2/\beta^2)B(\omega,v)v} 
=\left[1+O(\omega^2)\right]\exp\left[-(\beta\omega/2)^2(1-e^{-\tau})v\right],
\label{expan}
\end{equation}

Substituting Eq. (\ref{expan}) into Eq. (\ref{S_expan}) yields
\begin{equation}
S(z,v,\tau)\simeq\frac{2}{\pi}\int_0^\infty\frac{\sin\omega z}{\omega} 
\exp\left[-\frac{1}{4}\lambda(\tau,v)\omega^2\right]d\omega,
\label{SP_t0}
\end{equation}
where
\begin{equation}
\lambda(\tau,v)\equiv 2\theta\tau+2(1-e^{-\tau})v.
\label{lambda}
\end{equation}
The integral appearing in the right hand side of Eq. (\ref{SP_t0}) can be performed exactly \cite{erdelyi_it} and we get 
\begin{equation}
S(z,v,\tau)\simeq{\rm Erf}\left(\frac{z}{\sqrt{\lambda(v,\tau)}}\right), \qquad((\theta/\beta^2)\tau\gg 1),
\label{SP_t2}
\end{equation}
where 
$$
{\rm Erf}(x)=\frac{2}{\sqrt{\pi}}\int_0^x e^{-\xi^2}d\xi
$$
is the error function \cite{mos}. 

\begin{figure}[htb]
\begin{center}
\includegraphics{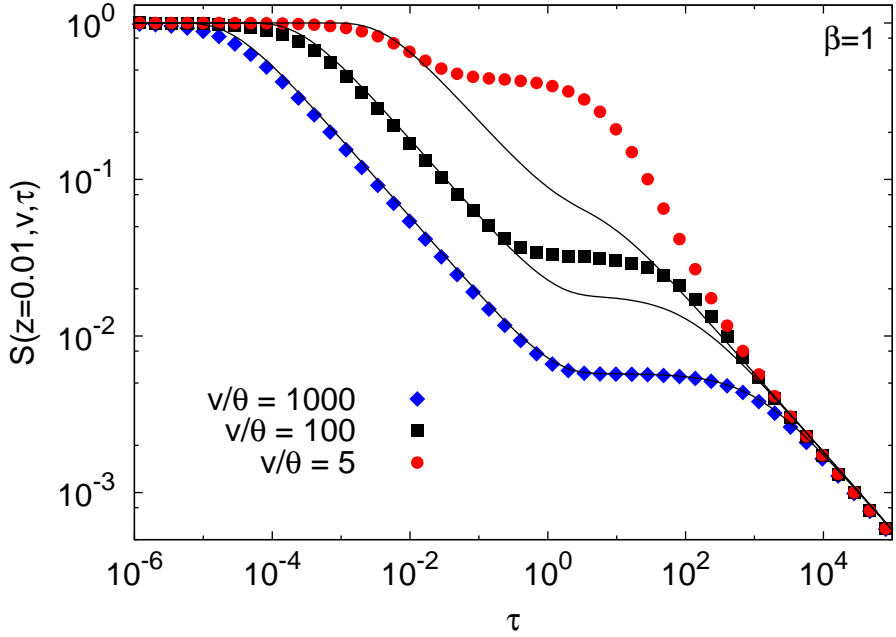}
\caption{(Color online) Log-log representation of the survival probability as a function of the scaled time $\tau$ with $z=0.01$ and $\beta=1$. Circles, squares and diamonds represent the numerical evaluation of the exact expression, Eq.~(\ref{exact_SP}), for the three values of the initial volatility $v$ shown in the figure. Solid lines correspond to the asymptotic expression given by Eq.~(\ref{SP_t2}) in terms of the error function and different curves correspond to different values of $v$. We see that as the volatility becomes smaller the exact expression needs a longer time to converge to the asymptotic expression. When the  initial volatility is five times larger than the normal level, the convergence time is about $10^5$ trading days. For $v/\theta=100$ this time is much less (about $10^3$ days). Let us note that for volatilities thousand times higher than the normal level, the asymptotic curve follows very closely the exact SP for all values of $\tau$ which is in agreement with the results of Sec. \ref{subsec32}. We use the realistic parameter values given in Fig. \ref{fig1}.}
\label{fig2}
\end{center}
\end{figure}

It is interesting to note that this approximation to the survival probability fulfills not only the boundary condition but the initial condition as well. The latter being quite remarkable since the approximation is based on the assumption of long time. That Eq. (\ref{SP_t2}) obeys the boundary condition is readily seen from the fact that ${\rm Erf}(0)=0$. As to the initial condition we have $S(z,v,\tau=0)={\rm Erf}(\infty)=1$. All of this is an indication of the goodness of the approximation as is corraborated by Fig.~\ref{fig2}.  

We shall now look at the mean first-passage time. Let us denote by $t_{FP}$ the time when the return $X(t)$, starting from a known value $x$ \cite{x=0}, attains a certain (critical) level $L$ for the {\it first time}. This time, usually referred to as the first-passage (or first-hitting) time, depends on $x$, $L$ and the current level of volatility $y$. It is evidently a random variable whose average $T=\langle t_{FP}\rangle$ is termed as the mean first-passage  time (MFPT). In terms of the survival probability this average hitting time is given by \cite{gardiner}
$$
T(x,y)=\int_0^\infty S(x,y,t)dt.
$$
Mathematical analysis tells us that in order for this integral to exist it is necessary that $S(x,y,t)$ decay faster than $1/t$ as $t\rightarrow\infty$. Now looking at Eq. (\ref{SP_t2}) and recalling that 
${\rm Erf(x)}=2x/\sqrt{\pi}+O(x^2)$ we see that the SP of the Heston model decays when $\tau\rightarrow\infty$ as
$$
S(z,v,\tau)\simeq\frac{2z}{\beta\sqrt{\pi\nu\tau}}\left[1+O\left(1/\tau\right)\right].
$$
Therefore, the survival probability falls off as $1/\sqrt{t}$ and the MFPT,
$$
T(x,y)=\infty,
$$
does not exist. 

\subsection{The first-passage problem for moderate and large volatilities}
\label{subsec32}

Let us start by observing, as it appears in the right hand side of Eq. (\ref{SP}), that the dependence of the SP on the volatility is that of a linear exponential. This suggests, as in the previous case when $(\theta/\beta^2)\tau\gg 1$, that as long as $v$ is not small --i.e., from moderate to large volatilities-- we may utilize again the saddle-point method for performing an approximate evaluation of Eq. (\ref{SP}). 

\begin{figure}
\includegraphics{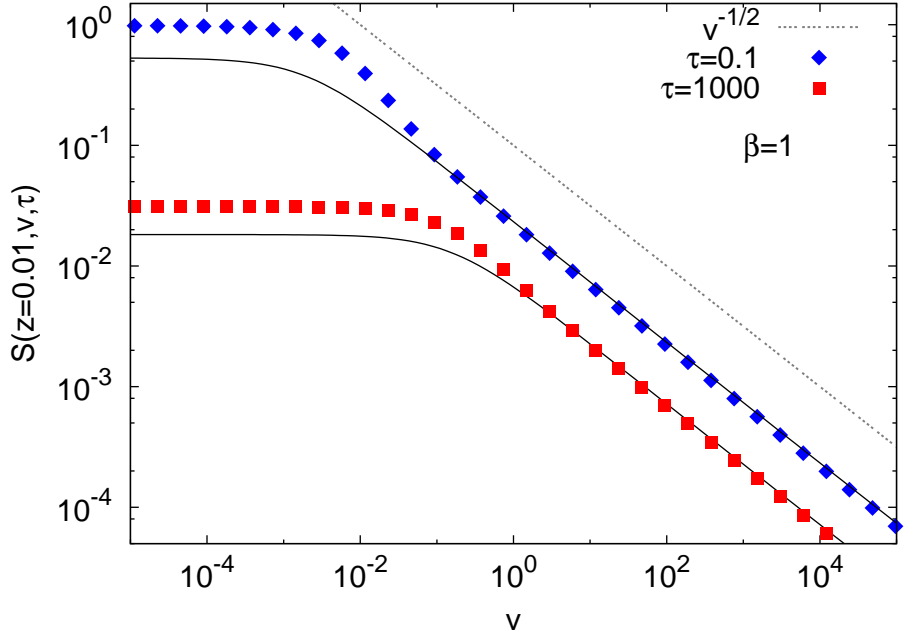}
\caption{(Color online) Log-log representation of the survival probability $S(z,v,\tau)$ as a function of the initial volatility with $z=0.01$, and $\beta=1$. Diamonds and squares correspond to the numerical evaluation of the exact expression~(\ref{exact_SP}) at two instants of time  $\tau=0.1$ and $\tau=1,000$. Solid lines represent the asymptotic error function given by Eq.~(\ref{SP_v2}). Note that the approximation improves as time increases, in agreement with the results of Sec. \ref{subsec31}. We also observe a power law decay of the form $v^{-1/2}$ with increasing volatility in agreement with the asymptotic estimate given by Eq.~(\ref{decreasing_SP}). We have taken the realistic parameters of Fig. \ref{fig1}.}
\label{fig3}
\end{figure}

The case of large volatility turns out to be completely analogous to the long time approximation just discussed. In fact both cases lead to exactly the same approximate expression for the SP. Indeed, from Eq. (\ref{b_expan}) we see that, like $\mu_-(\omega)$, $B(\tau,v)$ also attains a minimum value at $\omega=0$. Then, if $v$ is large enough, the saddle-point method allows us to write the exact SP, Eq. (\ref{SP}), in the approximate form
\begin{equation}
S(z,v,\tau)\simeq\frac{2}{\pi}\int_0^\infty \frac{\sin\omega z}{\omega}
\left[\frac{\Delta(\omega)e^{-\mu_-(\omega)\tau}}{\mu_+(\omega)+\mu_-(\omega)e^{-\Delta(\omega)\tau}}\right]^{2\theta/\beta^2}
e^{-\omega^2(1-e^{-\tau})v/2}d\omega.
\label{SP_v1}
\end{equation}
As before we expand to the lowest order in $\omega$ the bracketed expression in the integrand; we have (see Eqs. (\ref{minimum_mu}) and (\ref{delta_expan}))
$$
\frac{\Delta(\omega)e^{-\mu_-(\omega)\tau}}{\mu_+(\omega)+\mu_-(\omega)e^{-\Delta(\omega)(\omega)\tau}}\simeq
\frac{e^{-(\beta\omega/2)^2\tau}}{1+(\beta\omega/2)^2e^{-\tau}}=
e^{-(\beta\omega/2)^2\tau}\left[1+O\left(\omega^2\right)\right].
$$
Substituting this into Eq. (\ref{SP_v1}) we obtain again Eq. (\ref{SP_t0}), which proves that the approximate expression of the SP valid for moderate to large volatilities is also given by Eq. (\ref{SP_t2}):
\begin{equation}
S(z,v,\tau)\simeq{\rm Erf}\left(\frac{z}{\sqrt{\lambda(v,\tau)}}\right), \qquad(v \gg 1).
\label{SP_v2}
\end{equation}
The correctness of this approximation is clearly shown in Fig. \ref{fig3}. 

In the original units (cf. Eqs. (\ref{z}) and (\ref{tau-v})) the approximate SP reads
\begin{equation}
S(x,y,t)\simeq{\rm Erf}\left(\frac{|L-x|}{\sqrt{\lambda(t,y)}}\right), 
\label{SP_vol_dim}
\end{equation}
where (cf. Eq. (\ref{lambda}))
\begin{equation}
\lambda(t,y)=2m^2t+2(1-e^{-\alpha t})(y/\alpha).
\label{lambda_t}
\end{equation}
We remark that this approximation is valid for both long times and large volatilities. In the original units the 
condition of large volatilities, $v\gg 1$, is simply
$$
y\gg \alpha;
$$
while the condition of long times, $(\theta/\beta^2)\tau\gg 1$, reads
$$
t\gg \left(\frac{k}{\alpha m}\right)^2,
$$
that is, the time must be longer than a characteristic time, $t_c=(k/\alpha m)^2$, formed out of the three parameters of the Heston model: the vol of vol $k$, the reversion to the mean $\alpha$, and the normal level $m$. 

As long as we assume that neither the volatility nor the time are small, we can safely use approximation (\ref{SP_v2}) to obtain the behavior of the SP as $|L|\rightarrow\infty$. This is, perhaps, one of the most interesting aspect of the first-passage problem to look at, since the limit $|L|\rightarrow\infty$ is intimately related to extreme risks and, ultimately, to default. Using the following asymptotic expression of the error function \cite{mos}
$$
{\rm Erf}(x)\sim 1-\frac{e^{-x^2}}{\sqrt{\pi}x}\left[1+O(1/x^2)\right],
$$
we have
\begin{equation}
S(z,v,\tau)\sim 1-\frac{\sqrt{\lambda(v,\tau)}}{z\sqrt{\pi}} e^{-z^2/\lambda(v,\tau)}[1+O(1/z^2)].
\label{SP_large_L}
\end{equation}

The hitting probability to the level $L$, $W(z,v,\tau)$, is given by Eq. (\ref{default}). The default or uprising  probabilities will be obtained by assuming $|L|\rightarrow\infty$. In this way we find the following exponential decrease
\begin{equation}
W\sim \frac{e^{-L^2/\lambda}}{|L|} \qquad(|L|\rightarrow\infty).
\label{default_approx}
\end{equation}

We finally take a closer look at the behavior of the SP with increasing values of the volatility. 
Equation  (\ref{SP_v2}) is suitable for this purpose because it is valid for large values of the volatility. Intuition tells us that as $v\rightarrow\infty$ the survival probability should tend to zero which is readily seen from 
Eqs. (\ref{lambda}) and (\ref{SP_v2}). Let us elucidate how this limit works. We first note that
$$
\frac{z}{\sqrt{\lambda(\tau,v)}}=\frac{z}{\sqrt{2(1-e^{-\tau})v}}
\left[1+\frac{\nu\beta^2\tau}{4(1-e^{-\tau})v}+O\left(1/v^2\right)\right].
$$
Moreover \cite{mos}
$$
{\rm Erf}(x)=\frac{2}{\sqrt{\pi}}[x-x^3/3+O(x^5)].
$$
Merging these two expressions into Eq. (\ref{SP_v2}) we get
\begin{equation}
S(z,v,\tau)\simeq\frac{2z}{\sqrt{\pi(1-e^{-\tau})v}}
\left[1+O\left(1/v\right)\right].
\label{decreasing_SP}
\end{equation}
Therefore, when volatility $v$ increases the SP decreases as $v^{-1/2}$. Figure \ref{fig3} shows this power law behavior.

\section{The first-passage problem according to volatility fluctuations}
\label{sec4}

In the last section we have been dealing with approximate solutions to the hitting problem depending on the asymptotic values of time or volatility. We shall now look into other interesting and useful approximations. 

Recall that the parameter 
$$
\beta=k/\alpha
$$
measures the strength of the volatility fluctuations --given by the vol of vol $k$-- in relation to the deterministic pull --measured by $\alpha$-- toward the normal level of volatility. In this way we may term the dimensionless parameter $\beta$ as the ``normalized volatility fluctuation''. On the other hand from Eqs. (\ref{variance}) and (\ref{beta}) we see that $\beta^2=(2/\alpha m^2){\rm Var}\{Y(t)\}_{\rm st}$. In other words, $\beta^2$ is proportional to the amplitude of the volatility autocorrelation in the stationary state. Therefore, the higher $\beta$ is, the more intense the stationary autocorrelation becomes. We, therefore, expect two different patterns  according to whether $\beta$ is small or large. This duality is also present in the stationary distribution of the volatility. In effect, we see from Eq. (\ref{pst}) below that the behavior of $p_{st}(v)$ drastically changes as $\beta$ goes from small values bounded by $\beta<(2\theta)^{1/2}$ to larger values such that $\beta>(2\theta)^{1/2}$. Indeed, as $v\rightarrow 0$ the former case yields $p_{st}(v)\rightarrow 0$ while in the latter $p_{st}(v)\rightarrow\infty$; a fact that has to be taken into account in the numerical simulations of the Heston model \cite{moro}. 

We will explore which are the consequences of this on the first-passage problem of the return $X(t)$. Let us first investigate the case $\beta\rightarrow 0$. 

\subsection{Small fluctuations}
\label{subsec41}

We start form the exact expression of the SP given in Eq. (\ref{SP}) and suppose that volatility fluctuations are weak. Specifically, we assume that the vol of vol is much smaller than the pulling toward normal level, $k\ll\alpha$; that is
$$
\beta\ll 1.
$$
A glance at Eq. (\ref{delta}) suffices to realize that the assumption of $\beta\rightarrow 0$ is equivalent to assuming $\omega\rightarrow 0$. But the latter is precisely the limit taken in the previous section as a consequence of the saddle-point approximation. We therefore expect that the case of weak fluctuations will lead to the same approximation  than those of long times or large volatilities. Let us briefly show that this is indeed the case.

In effect, we start from Eq. (\ref{SP}) which we rewrite in the form
\begin{eqnarray}
S(z,v,\tau)=\frac{2}{\pi}\int_0^\infty &&\frac{\sin\omega z}{\omega}
\left[\frac{\Delta(\omega)}{\mu_+(\omega)+\mu_-(\omega)e^{-\Delta(\omega)\tau}}\right]^{2\theta/\beta^2}
\nonumber\\
&&\times\exp\left\{-\left(2/\beta^2\right)\left[\theta\mu_-(\omega)\tau+B(\omega,\tau)v\right]\right\}d\omega.
\label{SP_2}
\end{eqnarray}
If $\beta\rightarrow 0$ the integral can be approximately performed by the saddle-point method which implies the expansion of the exponential term around its maximum located as before at $\omega=0$. In other words:
$$
\exp\left\{-\left[\theta\mu_-(\omega)\tau+B(\omega,\tau)v\right]\right\}=
\exp\left\{\left[\theta\tau+\left(1-e^{-\tau}\right)v\right](\beta\omega/2)^2+O((\beta\omega)^4)\right\}.
$$
Moreover (cf. Eqs. (\ref{minimum_mu}) and (\ref{delta_expan}))
$$
\left[\frac{\Delta(\omega)}{\mu_+(\omega)+\mu_-(\omega)e^{-\Delta(\omega)\tau}}\right]^{2\theta/\beta^2}=1+O((\beta\omega)^2),
$$
and Eq. (\ref{SP_2}) yields
$$
S(z,v,\tau)\simeq\frac{2}{\pi}\int_0^\infty \frac{\sin\omega z}{\omega}
\exp\left\{-\left[\theta\tau+\left(1-e^{-\tau}\right)v\right](\omega^2/2)\right\}d\omega,
$$
which coincides exactly with Eq. (\ref{SP_t0}). Hence, we have again (see Eqs. (\ref{SP_v2})) 
\begin{equation}
S(z,v,\tau)\simeq{\rm Erf}\left(\frac{z}{\sqrt{\lambda(v,\tau)}}\right) \qquad(\beta\ll 1),
\label{SP_beta_small}
\end{equation}
with $\lambda(z,\tau)$ defined in Eq. (\ref{lambda}). In Fig. \ref{fig4} we show the goodness of this approximation for the small value $\beta=0.1$. 

Before proceeding further let us notice an apparent contradiction in the above statements on the coincidence of the approximate expressions for the SP between the cases of large volatility and reduced volatility fluctuations, because, at first sight, one would expect the opposite, i.e., that high volatility and enhanced volatility fluctuations would be equivalent. Let us remind, however, that a large volatility solely means that {\it today's volatility} is large and this, by any means, does not imply that volatility fluctuations must be intense and vice versa.

\subsection{Large fluctuations}
\label{subsec42}

\begin{figure}
\begin{center}
\includegraphics{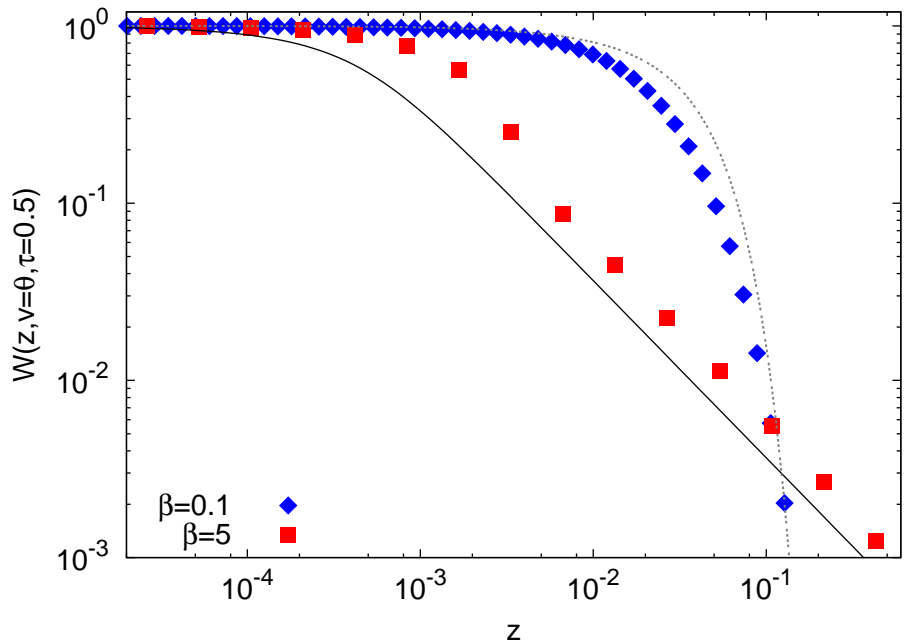}
\caption{(Color online) Log-log representation of the hitting probability $W(z,v,\tau)=1-S(z,v,\tau)$ as a function of the threshold distance $z$ at time $\tau=0.5$ and $v=\theta$. Diamonds and squares represent the numerical evaluation following the exact expression~(\ref{exact_SP}) for two different values of $\beta$. The dotted line corresponds to the asymptotic expression for $\beta$ small given by Eq.~(\ref{SP_beta_small}). The solid line represents the approximate expression obtained trhough Eq.~(\ref{SP_beta_large2}). We have chosen the realistic parameters of the previous figures and solely modify the parameter $k$ to provide different values of $\beta$.}
\label{fig4}
\end{center}
\end{figure}

We once more start with the exact expression of the SP given in Eq. (\ref{SP_2}) but supposing now intense volatility fluctuations, so that the vol of vol is greater than the pulling toward the normal level. In such a case
$$
\beta\gg 1,
$$
and from Eq. (\ref{delta}) we get the approximations
\begin{equation}
\Delta(\omega)=(\beta\omega)\left[1+O\left(\frac{1}{\beta^2}\right)\right], \qquad \mu_{\pm}(\omega)=(\beta\omega/2)\left[1\pm\frac{1}{\beta\omega}+O\left(\frac{1}{\beta^2}\right)\right].
\label{delta_large}
\end{equation}
Hence (cf. Eq. (\ref{B}))
\begin{equation}
B(\omega,\tau)\sim \beta\omega/2,
\label{B2}
\end{equation}
where, in writing Eq. (\ref{B2}), we have had to assume that time $\tau$ is long enough to neglect exponential terms of the form $e^{-\beta\omega\tau}$ and higher order terms. This certainly excludes the initial stages of the process as will become clearer later. Moreover
$$
\left[\frac{\Delta(\omega)}{\mu_+(\omega)+\mu_-(\omega)e^{-\Delta(\omega)\tau}}\right]^{2\theta/\beta^2}
\sim \left[1+O\left(\frac{1}{\beta^2}\right)\right],
$$
where we have proceeded as in Eq. (\ref{B2}) by neglecting exponentially small terms. Collecting these results into Eq. (\ref{SP_2}) we find
\begin{equation}
S(z,v,\tau)\sim\frac{2}{\pi}
\int_0^\infty \frac{\sin\omega z}{\omega}\exp\left[-\left(\theta\tau+v\right)(\omega/\beta)\right]d\omega,
\label{SP_beta_large1}
\end{equation}
which after performing the Fourier sine inversion yields \cite{erdelyi_it}
\begin{equation}
S(z,v,\tau)\sim\frac{2}{\pi}\arctan\left(\frac{\beta z}{\theta\tau+v}\right).
\label{SP_beta_large2}
\end{equation}
In Fig. \ref{fig4} we show the accuracy of this approximation for a large value of $\beta$.

In the original units we have  
\begin{equation}
S(x,y,t)\sim\frac{2}{\pi}\arctan\left(\frac{\beta|L-x|}{m^2t+y/\alpha}\right) \qquad(\beta\gg 1).
\label{SP_beta_large3}
\end{equation}
Let us remark that this expression verifies the boundary condition $S(L,y,t)=0$, but not the initial condition, since $S(x,y,0)\neq 1$ for $y\neq 0$. Consequently, Eq. (\ref{SP_beta_large3}) will work only after the initial period has elapsed, as otherwise was noted above. We observe, however, that the initial condition would be satisfied should we introduce an ad-hoc factor $(1-e^{\alpha t})$ multiplying the volatility. In this way we obtain a  ``semi-phenomenological approximation'', $S_p(x,y,t)$, which obeys the boundary condition as well as the initial condition:
\begin{equation}
S_p(x,y,t)=\frac{2}{\pi}\arctan\left(\frac{2|L-x|}{\lambda(t,y)}\right),
\label{pheno}
\end{equation}
where $\lambda(t,y)$ is defined in Eq. (\ref{lambda_t}). 

\section{Averaging the volatility}
\label{sec5}

Actual financial data consist in time series of prices from which returns are readily obtained. Once we have them, volatilities are constructed as the standard deviation of returns. Thus, in practice, volatility is a hidden variable that has to be measured by indirect means (we refer the reader to Ref. \cite{zoltan} for more information about this significant question). It is, therefore, meaningful to know whether the price of a given asset has reached for the first time some critical level regardless the value of its volatility. Let us remark that in physics an analogous situation would be knowing whether the position of a Brownian particle has first reached some preassigned value without worrying about its velocity \cite{masoliver_porra,masoliver_porra_2}.

Our goal is then to get the survival and hitting probabilities of the return overlooking volatility. We proceed as in Ref. \cite{met_heston} and average the volatility away from $S(z,v,\tau)$. In order to perform such an average we must choose an appropriate probability density for it. To this end we make the assumption that the entire process described by Eqs. (\ref{dx})-(\ref{dy}) commenced in the infinite past. Consequently, at the present time where we measure the return, the volatility has reached the stationary state. We, therefore, define the averaged SP, $S(z,\tau)$, as
\begin{equation}
S(z,\tau)=\int_0^\infty S(z,v,\tau)p_{st}(v)dv,
\label{averaged_SP}
\end{equation}
where $p_{st}(v)$, the stationary probability density of the volatility, is given by the Gamma distribution (cf. Eqs. (\ref{Gamma}), (\ref{theta}) and (\ref{beta})):
\begin{equation}
p_{st}(v)=\frac{(2/\beta^2)^{2\theta/\beta^2}}{\Gamma(2\theta/\beta^2)} v^{2\theta/\beta^2-1}e^{-2v/\beta^2}.
\label{pst}
\end{equation}

Substituting the exact SP, as given in Eq. (\ref{SP_2}), into Eq. (\ref{averaged_SP}) and performing the integral over the volatility we have
$$
S(z,\tau)=\frac{2}{\pi}\int_0^\infty\frac{\sin\omega z}{\omega}
\left[\frac{\Delta(\omega)}{(\mu_+(\omega)+\mu_-(\omega)e^{-\Delta(\omega)\tau})(1+B(\omega,\tau))}\right]^{2\theta/\beta^2}
e^{-(2\theta\tau/\beta^2)\mu_-(\omega)} d\omega.
$$
We will write this expression in a more convenient and simplified form. Let us note from Eq. (\ref{B}) that 
$$
1+B(\omega,\tau)=\frac{\mu_(\omega)+[1+\mu_-(\omega)]+\mu_-(\omega)[1-\mu_+(\omega)]e^{-\Delta(\omega)\tau}}
{\mu_+(\omega)+\mu_-(\omega)e^{-\Delta(\omega)\tau}},
$$
but $1+\mu_-(\omega)=\mu_+(\omega)$ and $1-\mu_+(\omega)=-\mu_-(\omega)$. Hence
$$
1+B(\omega,\tau)=
\frac{\mu_+^2(\omega)-\mu_-^2(\omega)e^{-\Delta(\omega)\tau}}{\mu_+(\omega)+\mu_-(\omega)e^{-\Delta(\omega)\tau}}.
$$
Therefore,
\begin{equation}
S(z,\tau)=\frac{2}{\pi}\int_0^\infty\frac{\sin\omega z}{\omega}
\left[\frac{\Delta(\omega)}{\mu_+^2(\omega)-\mu_-^2(\omega)e^{-\Delta(\omega)\tau}}\right]^{2\theta/\beta^2}
e^{-(2\theta\tau/\beta^2)\mu_-(\omega)} d\omega.
\label{SP_3}
\end{equation}
This is the exact expression for the return's survival probability when the volatility has been ``thermalized''. Before proceeding further in seeking approximate expressions for $S(z,\tau)$, as we have done above with $S(z,v,\tau)$, we will first obtain the SP had the return followed the Wiener process.

\subsection{The Wiener process}
\label{subsec51}

Undoubtedly the most spread market model is the geometric Brownian motion which was proposed by Osborne in the late ninety fifties \cite{osborne}. In this model the volatility $\sigma$ is constant and the zero-mean return is described by the stochastic differential equation
$$
dX(t)=\sigma dW(t),
$$
that is, $X(t)$ is the Wiener process with variance $\sigma^2$. 

Let us denote by $S_0(x,t)$ the survival probability of the Wiener process inside the semi-infinite strip $L<X(t)<\infty$ if $x>L$ (or inside $-\infty<X(t)<L$ if $x<L$). This function obeys the equation \cite{gardiner}
$$
\frac{\partial S_0}{\partial t}=\frac{1}{2}\sigma^2\frac{\partial^2 S_0}{\partial x^2},
$$
with initial and boundary conditions
$$
S_0(x,0)=1, \qquad S_0(L,t)=0.
$$

Getting along the lines of Sec. \ref{sec2} we define the new return variable (cf. Eq. (\ref{z})) 
$$
z=|L-x|
$$
for which the hitting problem remains unchanged, except that now the boundary is located at $z=0$. As before the solution is given in the form of a Fourier integral
$$
S_0(z,t)=\frac{2}{\pi}\int_0^\infty \tilde{S}_0(\omega,t)\sin\omega z d\omega,
$$
where $\tilde{S}_0(\omega,t)$ obeys the differential equation 
$$
\frac{\partial\tilde{S_0}}{\partial t}=-(\omega\sigma^2/2)\tilde{S}_0,
$$
with the initial condition
$$
\tilde{S}_0(\omega, 0)=\mathcal{P}(1/\omega).
$$

The solution to this initial-value problem is straightforward and reads
$$
\tilde{S}_0(\omega,t)=\mathcal{P}(1/\omega) e^{-\sigma^2\omega^2 t/2}.
$$
Hence
$$
S_0(z,t)=\frac{2}{\pi}\int_0^\infty e^{-\sigma^2\omega^2 t/2}\sin\omega z \frac{d\omega}{\omega},
$$
whence
\begin{equation}
S_0(x,t)={\rm Erf}\left(\frac{|L-x|}{\sqrt{2\sigma^2 t}}\right).
\label{SP_wiener}
\end{equation}

The comparison of the exact SP for the Wiener return, Eq. (\ref{SP_wiener}), with Heston's approximate SP given in Eq. (\ref{SP_vol_dim}) brings up an interesting detail. Thus for long times, large volatility or reduced volatility fluctuations, the approximate SP of the Heston model has the same form than that of Wiener SP, changing only the function $\sigma^2 t$ (which for the Wiener case is precisely the return's second moment) by the function $m^2 t+(1-e^{-\alpha t})(y/\alpha)$ for the Heston case. Let us observe that this function almost agrees with the return second moment which can be shown to be \cite{dragulescu}
\begin{equation}
\langle X^2(t)\rangle=m^2 t+(y-m^2)(1-e^{-\alpha t})/\alpha.
\label{second_moment}
\end{equation}

\subsection{Asymptotic approximations}
\label{subsec52}

We go back to the exact SP of Heston's return, Eq. (\ref{SP_3}). Looking at the exponential in the right hand side of that equation we easily realize that if the (dimensionless) group of parameters given by 
\begin{equation}
(\theta/\beta^2)\tau=(m/\beta)^2 t
\label{group}
\end{equation}
is large, we can apply the saddle-point approximation to get an approximate expression of the SP. Then proceeding as in Sec. \ref{subsec31} and using Eqs. (\ref{minimum_mu}) and (\ref{delta_expan}) we will have 
$$
S(z,\tau)\simeq \frac{2}{\pi}\int_0^\infty\frac{\sin\omega z}{\omega}e^{-(\theta\tau/2)\omega^2}d\omega.
$$
Whence
\begin{equation}
S(z,\tau)\simeq {\rm Erf}\left(\frac{z}{\sqrt{2\theta\tau}}\right).
\label{SP_3_approx(a)}
\end{equation}

\begin{figure}
\includegraphics{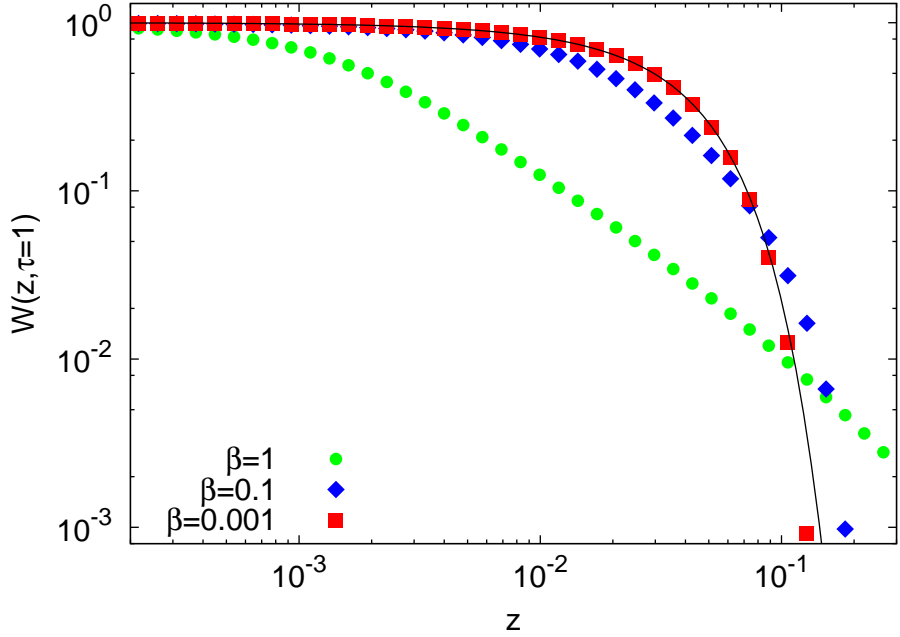}
\caption{(Color online) Log-log plot of the hitting probability $W(z,\tau)=1-S(z,\tau)$ as a function of $z$ for a fixed time ($\tau=1$, approximately $22$ trading days) and for different values of $\beta$. The solid line represents the approximation given by Eq. (\ref{SP_3_approx(a)}) --which, recall, is independent of $\beta$-- while dots (i.e., circles, diamonds and squares) represent the exact result. Equation (\ref{SP_3_approx(a)}) only works for very mild volatility fluctuations (note the clear divergence between the exact probability and its approximation for $\beta=1$). Parameters are the same than those of Fig. \ref{fig1}.}
\label{fig5}
\end{figure}

\begin{figure}
\includegraphics{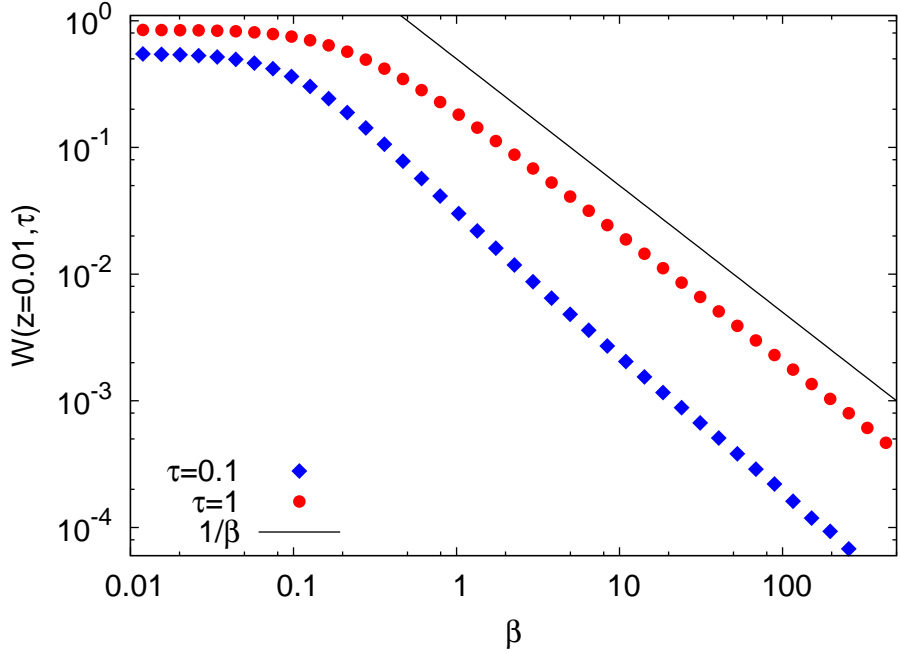}
\caption{(Color online) Log-log plot of the hitting probability $W(z,\tau)=1-S(z,\tau)$ as a function of $\beta$. Circles and diamonds represent the exact result, obtained through numerical integration of Eq.~(\ref{SP_3}), for $z=0.01$ and two instants of time: $\tau=0.1$ and $\tau=1$. Note that for small values of $\beta$ the hitting probability is insensitive to $\beta$, in agreement with Eq. (\ref{SP_3_approx(a)}). Moreover, for large values of $\beta$ we observe the power-law decay predicted in the asymptotic estimate given by Eq. (\ref{L_default}), i.e., $W\sim\beta^{-1}$. Parameters are the same than those of Fig. \ref{fig1}.}
\label{fig6}
\end{figure}

Figure \ref{fig5} shows the hitting probability, $W(z,\tau)=1-S(z,\tau)$, with $S(z,\tau)$ obtained by means of Eq. (\ref{SP_3_approx(a)}). We compare it with the exact result evaluated through Eq. (\ref{SP_3}). As we see in Fig. \ref{fig5} the approximation works fairly well for very small values of $\beta$, whereas for $\beta=1$ Eq. (\ref{SP_3_approx(a)}) significantly deviates from the exact result. We remark that although the approximation above is meant for small values of $\beta$ it is itself independent of $\beta$. In Fig. \ref{fig6} we represent the dependence on $\beta$ of the hitting probability and this fact becomes apparent.

In the original units we have
\begin{equation}
S(x,t)\simeq {\rm Erf}\left(\frac{|L-x|}{\sqrt{2m^2 t}}\right).
\label{SP_3_approx}
\end{equation}
Recall that this approximation is valid as long as the group of parameters defined in Eq. (\ref{group}),
$$
\left(m/\beta\right)^2 t\gg 1,
$$
attains a large value. That is to say, either for long time $t$, or when the normal level $m$ is large, or the (normalized) volatility fluctuation $\beta$ is small. Moreover Eq. (\ref{SP_3_approx}) exactly corresponds to the Wiener SP, Eq. (\ref{SP_wiener}), by switching the constant volatility of the latter to the normal level. We also note this approximation has the same form than that of Eq. (\ref{SP_beta_small}) after setting $y=0$.   

Since the approximate expression for $S(x,t)$ given in Eq. (\ref{SP_3_approx(a)}) is the same than that of $S(z,v=0,\tau)$ given in Eq. (\ref{SP_beta_small}), the default (or uprising) problem when $|L|\rightarrow\infty$ will be given by Eqs. (\ref{SP_large_L}) and (\ref{default_approx}) when $v=0$. That is,
\begin{equation}
S(z,\tau)\sim 1-\sqrt{\frac{2\theta\tau}{\pi}}\frac{e^{-L^2/2\theta\tau´}}{|L|} \qquad(|L|\rightarrow\infty),
\label{SP_3_large_L}
\end{equation}
and
\begin{equation}
W(z,\tau)\sim \sqrt{\frac{2\theta\tau}{\pi}}\frac{e^{-L^2/2\theta\tau´}}{|L|} \qquad(|L|\rightarrow\infty),
\label{default_3_approx}
\end{equation}
which means an acute exponential decrease of default as $|L|\rightarrow\infty$. This is to be contrasted with a very much slow descent of these probabilities to be discussed next.  

\subsection{High volatility fluctuations}
\label{subsec53}

The other approximation that remains is that of high volatility fluctuations. We start from the exact expression of the SP given in Eq. (\ref{SP_3}) and if $\beta$ is large we may proceed as in Sec. \ref{subsec41}. We, therefore, use the expansions given in Eq. (\ref{delta_large}) along with
$$
\mu_{\pm}^2(\omega)=(\beta\omega/2)^2\left[1\pm\frac{2}{\beta\omega}+O\left(\frac{1}{\beta^2}\right)\right],
$$
to write
$$
\frac{\Delta(\omega)}{\mu_+^2(\omega)-\mu_-^2(\omega)e^{-\Delta(\omega)\tau}}\sim\frac{2}{1+\beta\omega/2},
$$
where we have neglected exponential small terms of the form $e^{-\beta\omega\tau}$. Substituting into Eq. (\ref{SP_3}) we obtain
$$
S(z,\tau)\sim\frac{2}{\pi}\int_0^\infty\frac{\sin\omega z}{\omega}
\left[\frac{2}{1+\beta\omega/2}\right]^{2\theta/\beta^2}
e^{-(\theta\tau/\beta)\omega} d\omega.
$$

Moreover as $\beta\rightarrow\infty$ we have
$$
\left[\frac{2}{1+\beta\omega/2}\right]^{2\theta/\beta^2}=1+
\left(\frac{2\theta}{\beta^2}\right)\ln\left[\frac{2}{1+\beta\omega/2}\right]+O\left(\frac{1}{\beta^4}\right),
$$
and to the lowest order we get
$$
S(z,\tau)\sim\frac{2}{\pi}\int_0^\infty\frac{\sin\omega z}{\omega}
\left[1+O(1/\beta^2)\right]e^{-(\theta\tau/\beta)\omega} d\omega.
$$
Whence
\begin{equation}
S(z,\tau)\sim\frac{2}{\pi}\arctan\left(\frac{\beta z}{\theta\tau}\right), \qquad(\beta\gg 1).
\label{SP_final}
\end{equation}

\begin{figure}
\includegraphics{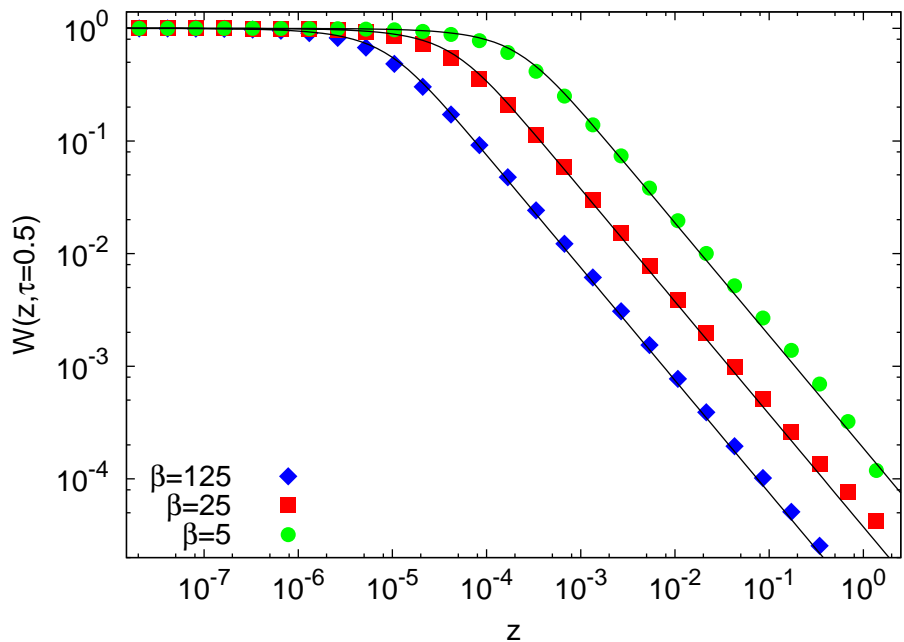}
\caption{(Color online) Log-log plot of the hitting probability $W(z,\tau)=1-S(z,\tau)$ as a function of $z$ with $\tau=0.5$ and for three large values of $\beta$. Solid lines display $W(z,\tau)$ based on the approximation given in Eq. (\ref{SP_final}) while dots represent the exact result obtained through the numerical integration of Eq.~(\ref{SP_3}). Parameters are the same than those of Fig. \ref{fig1}.}
\label{fig7}
\end{figure}

The asymptotic SP given by Eq. (\ref{SP_final}) has the same functional form than that of Eq. (\ref{SP_beta_large2}). In fact by neglecting the volatility in Eq. (\ref{SP_beta_large2}) both approximations coincide. A singular characteristic of Eq. (\ref{SP_final}) is that it satisfies both boundary and initial conditions, while the approximation (\ref{SP_beta_large2}) involving volatility solely obeys the boundary condition. Moreover, and contrary to the approximation given by Eq. (\ref{SP_3_approx}), the case of intense volatility fluctuations given in Eq. (\ref{SP_final}) is not by any means related to the Wiener case discussed in Sec \ref{subsec51}. Therefore,  Eq. (\ref{SP_final}), and the same applies to Eq. (\ref{SP_beta_large2}), represents a distinct characteristic of The Heston model --and, by extension, to any SV model-- which has no parallel in the Wiener model.

Figure \ref{fig7} displays the hitting probability $W(z,\tau)$ when the SP is given by approximation (\ref{SP_final}) along with the exact $W$ obtained through the numerical integration of Eq. (\ref{SP_3}). The curves shown correspond to increasingly large values of $\beta$. Let us note the excellent agreement between the approximation and the exact result even for moderately high values of $\beta$.  

The singular character of Eq. (\ref{SP_final}) is best expressed when one considers the possibility of default (or uprising) for which $|L|\rightarrow\infty$. In this case, using the asymptotic estimate 
$$
\arctan z\sim \pi/2-1/z+O(1/z^3),
$$
we get
\begin{equation}
S(z,\tau)\sim 1-\frac{\theta\tau}{\beta|L|} \qquad(|L|\rightarrow\infty),
\label{L_final}
\end{equation}
and
\begin{equation}
W(z,\tau)\sim \frac{\theta\tau}{\beta|L|} \qquad(|L|\rightarrow\infty),
\label{L_default}
\end{equation}
(see Fig. \ref{fig6}). Comparing this slow decrease in $L$ with the enhanced exponential fall of the previous section (see Eqs. (\ref{SP_3_large_L})-(\ref{default_3_approx})), we see that a boost on volatility fluctuations results in a large increase of risk. 

\begin{figure}
\includegraphics{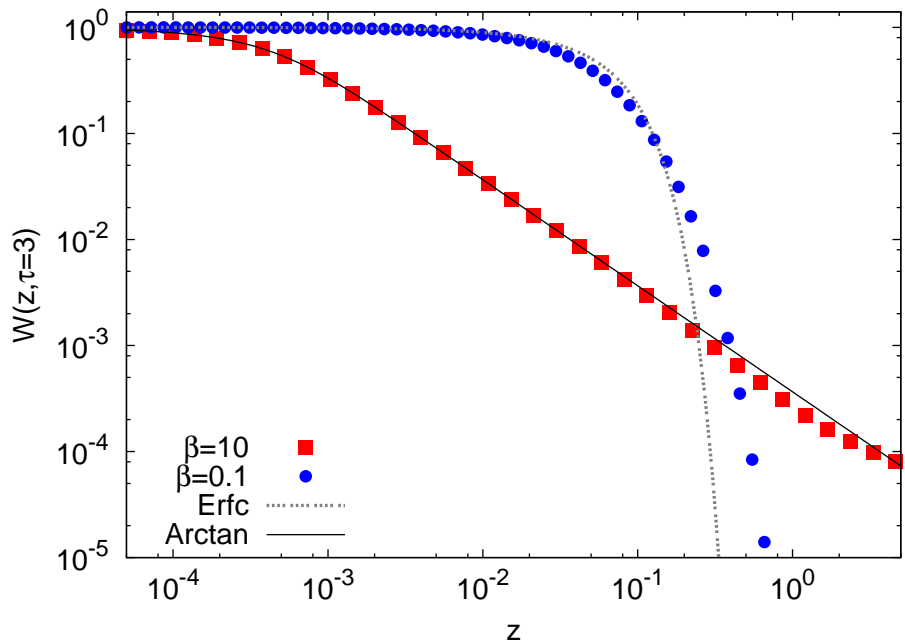}
\caption{(Color online) Log-log plot of the hitting probability $W(z,\tau)=1-S(z,\tau)$ as a function of $z$ with $\tau=3$ and for large and small volatility fluctuations given, respectively, by $\beta=10$ and $\beta=0.1$. Circles and squares represent the exact $W(z,\tau)$ obtained through Eq. (\ref{SP_3}). Curves display $W(z,\tau)$ based on the approximations given in Eqs. (\ref{SP_3_approx(a)}) (dotted line) and (\ref{SP_final}) (solid line). Parameters are the same than those of Fig. \ref{fig1}}
\label{fig8}
\end{figure}

\subsection{The crossing level and increase of risk}
\label{subsec54}

Figure \ref{fig8} shows the exact hitting probability for both mild and intense volatility fluctuations, along with the approximations given by Eqs. (\ref{SP_3_approx(a)}) and (\ref{SP_final}) corresponding, respectively, to $\beta$ small and large. Considering $z=|L|$ \cite{x=0} the exact expressions of $W$ intersect at certain value $l_c$ of $|L|$ (in Fig. \ref{fig8}, $l_c\simeq 0.336$). The intersection marks a turning point of the hitting problem, for when $|L|<l_c$ the probability of reaching $|L|$ is higher for small volatility fluctuations than when these are intense. On the other hand, when $|L|>l_c$ the situation is reversed, because now the hitting probability corresponding to large values of $\beta$ becomes increasingly higher with $|L|$ than that corresponding to small values of $\beta$. All of this is in agreement with the asymptotic estimates given by Eqs. (\ref{default_3_approx}) and (\ref{L_default}) which predict a quadratic exponential decay in $|L|$ for $\beta$ small, and a much slower power-law decay of the form $1/|L|$ when $\beta$ is large. 

\begin{figure}
\includegraphics{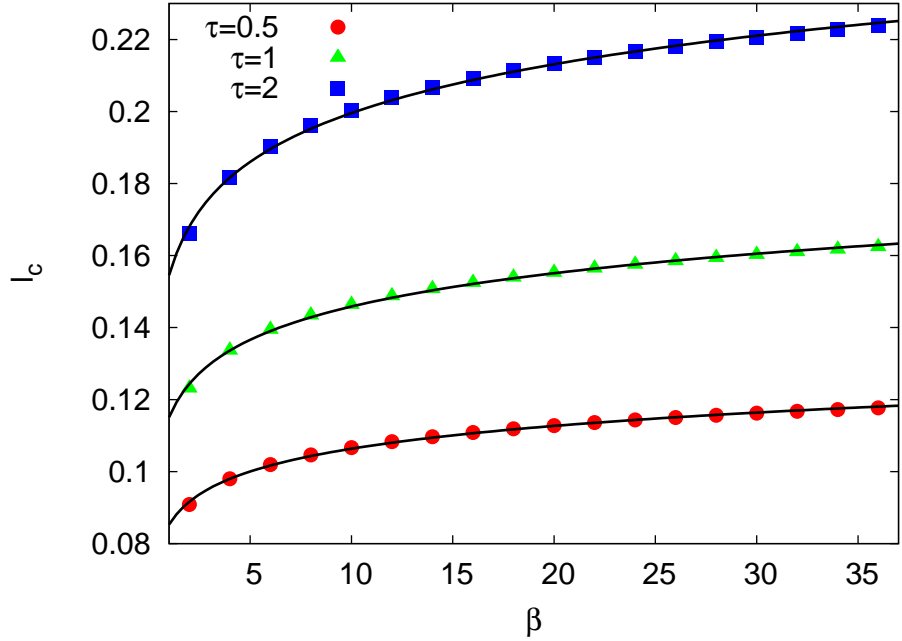}
\caption{(Color online) Numerical solution of Eq. (\ref{lc0}) in terms of $\beta$ with a fixed value of the normal level of volatility $\theta=1.92\times 10^{-3}$ (the same of the previous figures) and for $\tau=0.7$, $\tau=1.3$ and $\tau=2.0$. Note that for approximately $\beta>10$, $l_c$ slows down its increase. Solid lines represent a logarithmic fit. Parameters are the same than those of Fig. \ref{fig1}.}
\label{fig9}
\end{figure}

We can approximately evaluate the crossing level $l_c$ through the asymptotic forms of the SP given in Eqs. (\ref{SP_3_approx(a)}) and (\ref{SP_final}). Indeed, due to the good precision of these approximations (see Fig. \ref{fig8}), a relatively accurate estimation of $l_c$ is obtained from the solution of the transcendental equation 
\begin{equation}
{\rm Erf}\left(\frac{l_c}{\sqrt{2\theta\tau}}\right)=\frac{2}{\pi}\arctan\left(\frac{\beta l_c}{\theta\tau}\right),
\label{lc0}
\end{equation}
which, in turn, shows that the crossing level depends on volatility fluctuations, $\beta$, and also on time and the normal level through the combination $\theta\tau$, i.e., 
$$
l_c=l_c(\beta,\theta\tau).
$$
In Fig. \ref{fig9} we represent the numerical solution of Eq. (\ref{lc0}) in terms of $\beta$, with $\theta$ fixed and for three instants of time. We see there that $l_c$ increases with time and also with $\beta$. In fact, there seems to be a logarithmic behavior of the crossing level with respect to the volatility fluctuations:
\begin{equation}
l_c\sim\log\beta,
\label{lc1}
\end{equation}
as is shown by the fits in Fig. \ref{fig9}. 

\begin{figure}
\includegraphics{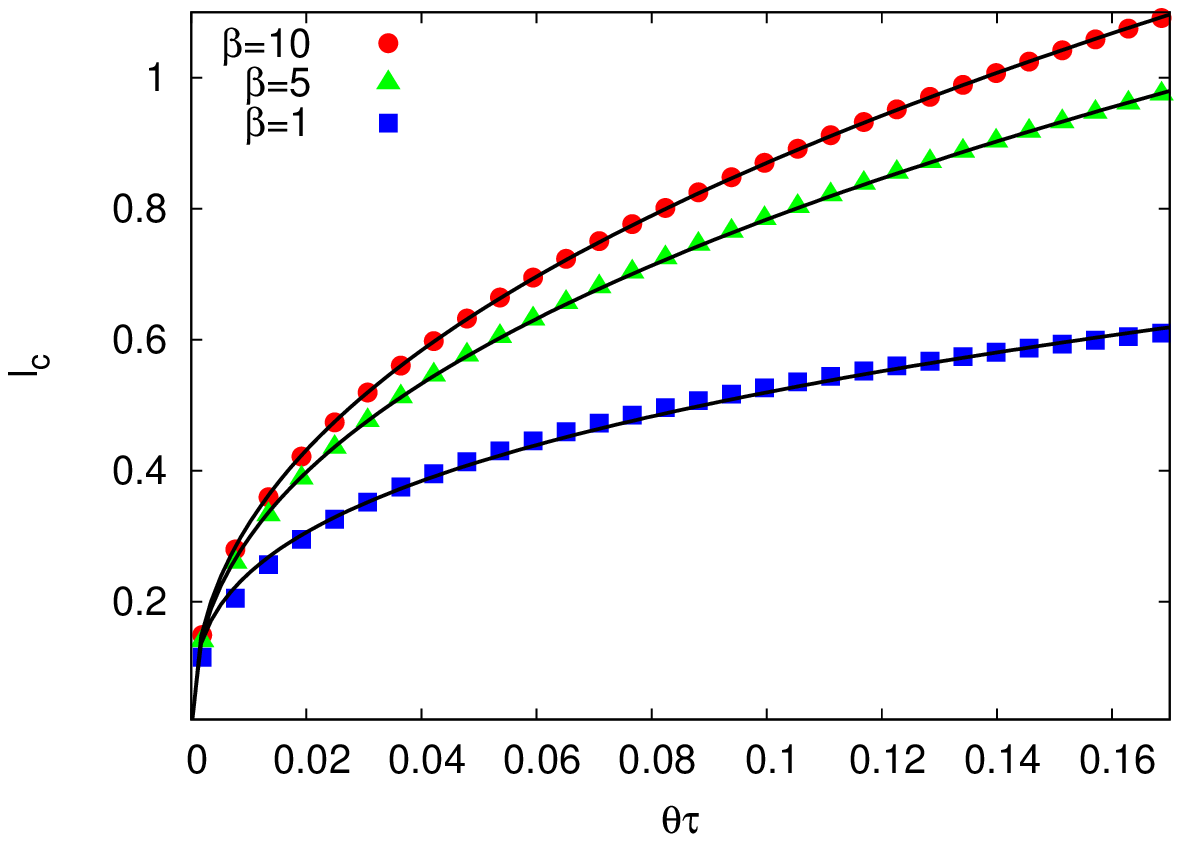}
\caption{(Color online) Numerical solution of Eq. (\ref{lc0}) in terms of $\theta\tau$ for $\beta=1$, $\beta=5$ and $\beta=10$. Solid lines are the power law given in Eq. (\ref{lc2}) where the exponent $\gamma$ depends on $\beta$ and takes the values $\gamma=0.3293\pm 0.0051$ for $\beta=1$, $\gamma=0.4212\pm 0.0011$ for $\beta=5$ and $\gamma=0.4358\pm 0.0007$ for $\beta=10$. Parameters are the same than those of Fig. \ref{fig1}.}
\label{fig9b}
\end{figure}

The dependence on time of the crossing level is pictured in Fig. \ref{fig9b} where we clearly see that $l_c$ increases with time, although now there seems to be a power-law behavior
\begin{equation}
l_c\sim(\theta\tau)^\gamma,
\label{lc2}
\end{equation}
where $\gamma$ depends on $\beta$.

Let us finally return to the the increase of risk as volatility fluctuations get higher. This can be visualized by calculating the ratio
\begin{equation}
\bar{W}(z,\tau)\equiv\frac{W(z,\tau)}{W_0(z,\tau)},
\label{ratio}
\end{equation}
between the hitting probability $W(z,\tau)$ (evaluated through the exact SP, Eq. (\ref{SP_3})) and the hitting probability corresponding to the Wiener process $W_0(z,\tau)=1-S_0(z,\tau)$, where $S_0(z,\tau)$ is given in Eq. (\ref{SP_wiener}). Note that we use $W_0(z,\tau)$ to construct the ratio $\bar{W}(z,\tau)$ because for small volatility fluctuations Heston's and Wiener's SPs share the same form.

\begin{figure}
\includegraphics{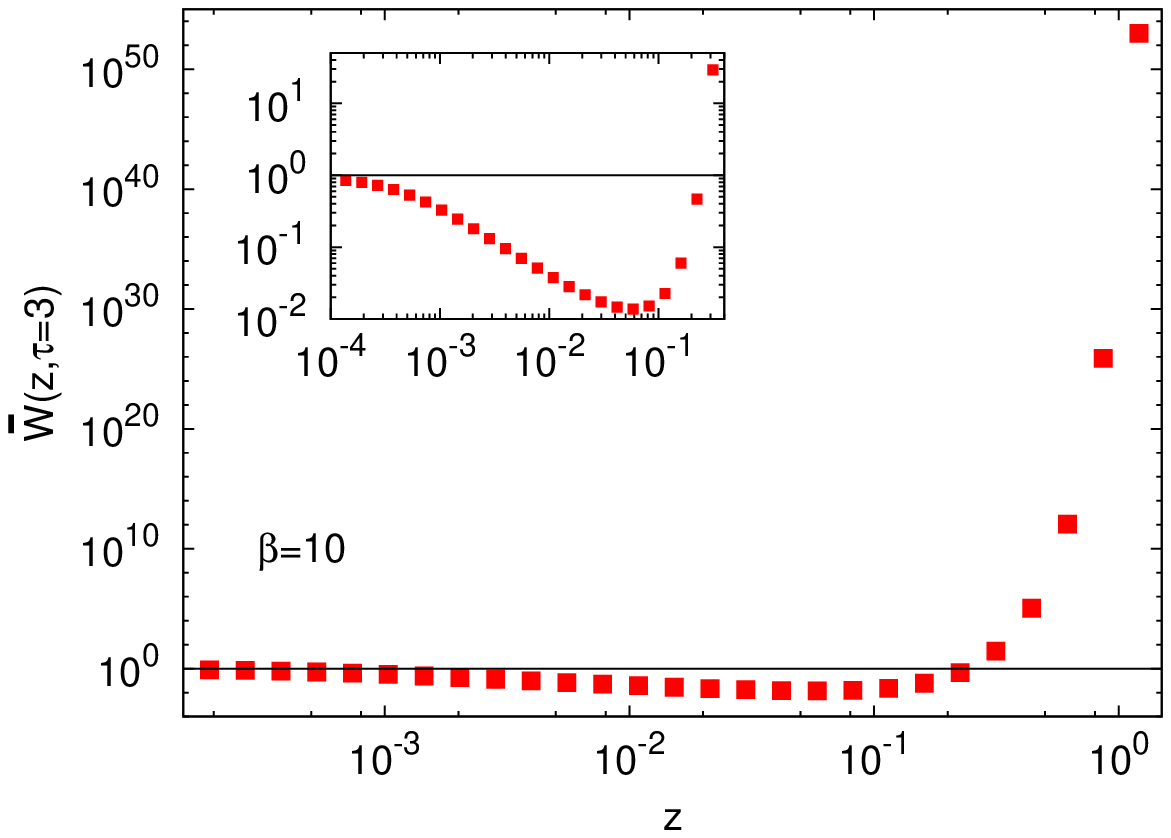}
\caption{(Color online) Log-log plot of the ratio $\bar{W}(z,\tau)$ as a function of $z$ when $\beta=10$ and $\tau=3$. Note the sharp burst as $|L|$ increases, in agreement with the approximation given by Eq. (\ref{ratio_1}). The inset shows the little  decrease of the ratio for small values of $|L|$ (see also Fig. \ref{fig8}). The turning point between these two behaviors is approximately located at $|L|=l_c=0.336$.}
\label{fig10}
\end{figure}

For sufficiently large fluctuations, and according to the above discussion, the ratio $\bar{W}$ should greatly increase as $|L|\rightarrow\infty$. Indeed, when $\beta\gg 1$ and $|L|\rightarrow\infty$ we have from Eqs. (\ref{default_3_approx}) and (\ref{L_default}) 
\begin{equation}
\bar{W}(z,\tau)\sim(\pi\theta\tau/2)^{1/2}e^{L^2/2\theta\tau}.
\label{ratio_1}
\end{equation}
In Fig. \ref{fig10} we represent $\bar{W}(z,\tau)$ (evaluated from the exact result) as a function of $z$ (i.e., $|L|$) when $\beta=10$ and $\tau=3$. As we see there the exponential growth of $\bar{W}$ as $|L|$ increases is manifest. 

Figure \ref{fig10} also shows that for small values of $|L|$ the ratio is smaller than $1$, which implies a greater probability of hitting label $|L|$ when $\beta$ is small than when it is large. However, this fact, already observed in Fig. \ref{fig8}, has little consequences on risk control due to the smallness of $|L|$.  

We consider the above findings on the relationship between volatility fluctuations and risk one of the key results of the present work which might have useful practical consequences on risk management and control.

\section{Summary and Conclusions}
\label{sec6}

We have approached the issue of risk evaluation and control as a first-passage problem within the assumption of stochastic volatility given by the Heston model. The problem is solved when one knows the survival probability,  $S(x,y,t)$, to a certain critical level $L$, of the return $X(t)$ starting at $X(0)=x$ (usually $x=0$ \cite{x=0}) with initial volatility $Y(0)=y$. Closely related to the SP it is the hitting probability $W(x,y,t)=1-S(x,y,t)$. When the critical level, $L>0$, is positive $W$ represents the probability of an uprising at time $t$. On the contrary if $L<0$, $W$ gives the probability of having, at time $t$, a loss quantified by $L$ (when $L\rightarrow-\infty$ this is the default probability). 

For the Heston model, the SP obeys the backward Fokker-Planck equation, Eq. (\ref{fpe}), with initial and boundary conditions given by Eq. (\ref{ibc}). The problem have been solved exactly by means of the Fourier-sine integral given in Eq. (\ref{SP}). We have then proceeded to get handy approximations to the exact SP. Thus, in the asymptotic regimes of either long times or large initial volatilities, the SP goes to the following Gaussian form which constitutes a restatement of the central limit theorem:
\begin{equation}
S(x,y,t)\simeq{\rm Erf}\left(\frac{|L-x|}{\sqrt{\lambda(t,y)}}\right),
\label{gaussian_SP}
\end{equation}
where $\lambda(t,y)$ is defined in Eq. (\ref{lambda_t}). This approximation is also valid when the amplitude of the volatility fluctuations is small. This amplitude is characterized by the dimensionless parameter $\beta=k/\alpha$ 
formed by the ratio between the vol-of-vol $k$ and the strength of the reverting force, toward the normal level, measured by $\alpha$. A remarkable feature of the asymptotic expression, Eq. (\ref{gaussian_SP}), is that it has the same form had the return followed the Wiener process (i.e., constant volatility) instead of 
the Heston model.

Other consequences of the asymptotic form above are, for one hand, the non-existence of a mean first-passage time, since 
$$
S(x,y,t)\sim t^{-1/2}
$$
as $t\rightarrow\infty$ and the SP does not decays fast enough in order to possess a mean first-passage time. On the other hand, when the volatility variable $y$ increases the SP decreases as the power law:
$$
S(x,y,t)\sim y^{-1/2}.
$$
Obviously these properties are shared by the Wiener process because all of them stem from the common form given by Eq. (\ref{gaussian_SP}). 

An additional asymptotic form of the survival probability is obtained when one considers the fluctuations of the volatility which are characterized by the parameter $\beta$. As mentioned above, for mild fluctuations for which $\beta\ll 1$, the approximate SP is still given by Eq. (\ref{gaussian_SP}). However, when $\beta\gg 1$, i.e., for extreme fluctuations, the asymptotic SP is given by the following non-Gaussian form:
\begin{equation}
S(x,y,t)\sim\frac{2}{\pi}\arctan\left(\frac{\beta|L-x|}{m^2t+y/\alpha}\right) \qquad(\beta\gg 1).
\label{non_Gaussian_SP}
\end{equation}

Real financial data consist of time series of prices and the volatility is not directly recorded and only observed in an indirect way. This hidden character makes it worth averaging out volatility from the expressions of $S(x,y,t)$ and thus solving the hitting problem for the return alone. The assumption to be made is that the volatility has reached the stationary state characterized by the Gamma distribution. 

Following this way we have obtained the exact expression of $S(x,t)$ given in Eq. (\ref{SP_3}), again in terms of a Fourier integral. As before the averaged SP has two different asymptotic forms: a Gaussian approximation for long times or small volatility fluctuations and another form, which is non-Gaussian, for intense fluctuations. Thus, when the group of parameters $(m/\beta)^2t\gg 1$ attains a large value (i.e., for long time $t$, or large normal level $m^2$, or small volatility fluctuations $\beta$) the averaged SP is approximately given by the following Gaussian form
\begin{equation}
S(x,t)\simeq {\rm Erf}\left(\frac{|L-x|}{\sqrt{2m^2t}}\right),
\label{gaussian_SP_x}
\end{equation}
which, curiously enough, has the same form than Eq. (\ref{gaussian_SP}) after setting $y=0$. 

The same trend is obtained in the case of large volatility fluctuations, since now the following non-Gaussian form emerges
\begin{equation}
S(x,t)\sim\frac{2}{\pi}\arctan\left(\frac{\beta|L-x|}{m^2t}\right) \qquad(\beta\gg 1),
\label{non_Gaussian_SP_x}
\end{equation}
which, again, has the same form of Eq. (\ref{non_Gaussian_SP}) after setting $y=0$. 

The most striking difference between these two asymptotic forms of the first-passage problem appears when considering the extreme risk of default. In such a case $L\rightarrow-\infty$ and the hitting probability, $W(x,t)$, corresponding to the Gaussian form (\ref{gaussian_SP_x}) quickly decays by following a decreasing quadratic exponential. On the other hand, as volatility fluctuations increase, the non-Gaussian approximation, Eq. (\ref{non_Gaussian_SP_x}), decays much slowly by following a power law. This means a significant increase of risk as the fluctuations of the volatility soar. We have shown that there is a crossing level $l_c=l_c(\beta,m^2t)$ from which risk exponentially rises. For a fixed value of time, the crossing level seems to follow the logarithmic law $l_c\sim \log\beta$, while for a fixed value of $\beta$ the crossing level apparently increases as a power law $l_c\sim(m^2t)^\gamma$, where the exponent $\gamma$ depends of $\beta$.

Practical consequences of these findings on actual markets will be presented in forthcoming works.

\acknowledgments 
Partial financial support from Direcci\'on General de Investigaci\'on under contract No. FIS2006-05204 is acknowledged.

\appendix

\section{Functions $A(\omega,\tau)$ and $B(\omega,\tau)$}
\label{appB}

Function $B(\omega,\tau)$ is the solution of the Riccati equation
\begin{equation}
\dot{B}=-B-B^2+(\beta\omega/2)^2,
\label{b1}
\end{equation}
with initial condition $B(\omega,0)=0$. Define a new function $Z(\omega,\tau)$ related to $B$ by
$$
B=\frac{\dot{Z}}{Z},
$$
then $Z$ obeys the linear equation
$$
\ddot{Z}+\dot{Z}-(\beta\omega/2)^2Z=0,
$$
whose solution is
$$
Z(\omega,\tau)=C_1(\omega)e^{\mu_-(\omega)\tau}+C_2(\omega)e^{-\mu_+(\omega)\tau},
$$
where $C_1(\omega)$ and $C_2(\omega)$ are arbitrary and $\mu_\pm(\omega)$ are defined in Eq. (\ref{delta}). 
The expression of $B(\omega,\tau)$ is thus given by
$$
B(\omega,\tau)=\frac{\mu_-(\omega)-[C_2(\omega)/C_1(\omega)]\mu_+(\omega)e^{-\Delta(\omega)\tau}}
{1+[C_2(\omega)/C_1(\omega)]e^{-\Delta(\omega)\tau}},
$$
where $\Delta(\omega)$ is defined in Eq. (\ref{delta}). From the initial condition we get
$$
C_2(\omega)/C_1(\omega)=\mu_-(\omega)/\mu_+(\omega).
$$
Therefore, 
\begin{equation}
B(\omega,\tau)=\mu_-(\omega)
\frac{1-e^{-\Delta(\omega)\tau}}{1+[\mu_-(\omega)/\mu_+(\omega)]e^{-\Delta(\omega)\tau}},
\label{b2}
\end{equation}
which proves Eq. (\ref{B}).

We now substitute Eq. (\ref{b2}) into Eq. (\ref{A0}) and define $\zeta=e^{-\Delta(\omega)}$ as a new integration variable, we get
$$
A(\omega,\tau)=\left(\frac{2\theta}{\beta^2}\right)\frac{\mu_-(\omega)}{\Delta(\omega)}\int_{e^{-\Delta(\omega)\tau}}^1
\frac{1-\zeta}{\zeta[1+(\mu_-/\mu_+)\zeta]}d\zeta.
$$
Taking into account
$$
\int \frac{1-\zeta}{\zeta[1+(\mu_-/\mu_+)\zeta]}d\zeta=
\ln\zeta-(1+\mu_+/\mu_-)\ln[1+(\mu_+/\mu_-)\zeta],
$$
and recalling that $\mu_-(\omega)+\mu_+(\omega)=\Delta(\omega)$ (cf. Eq. (\ref{delta})), we finally obtain 
\begin{equation}
A(\omega,\tau)=\left(\frac{2\theta}{\beta^2}\right)\left\{\mu_-(\omega)\tau+\ln\left[\frac{\mu_+(\omega)+\mu_-(\omega)e^{-\Delta(\omega)\tau}}{\Delta(\omega)}\right]\right\},
\label{b3}
\end{equation}
which proves Eq. (\ref{A}).


\begin{thebibliography}{99}
\bibitem{footnote1} We do not consider here the possible existence of cross correlation between the input noises $\xi_1(t)$ and $\xi_2(t)$. We will explore the influence on the escape problem of a possible ``leverage'' correlation in a future work. Ref.~\cite{bonano1} has addressed this issue in a modified Heston process.
\bibitem{x=0} Usually $x$ is today's return and, by definition, $x=0$. However, to avoid confusion we keep the symbol $x$.
\end{thebibliography}
\end{document}